\documentstyle[12pt]{article}
\begin{document}

\baselineskip 20pt 

\vspace{.5in}

\begin{center}

{\Large \bf   
Electronic states\\
in\\ideal free standing films
}
\vskip .8in
{\bf
Shang Yuan Ren\\
Department of Physics, Peking University \\
Beijing 100871, People's Republic of China\\
}
\vbox{}
\end{center}
\newpage
\begin{center}
{\bf \large Abstract}
\end{center}
\par
Exact and general results on 
the electronic states in ideal free standing films bounded between
$\tau_3 {\bf a}_3$ and $(\tau_3 + N_3) {\bf a}_3$ - here ${\bf a}_3$ is
the only primitive vector out of the film plane, $\tau_3$ is the bottom
boundary and $N_3$ is a positive integer indicating the thickness of
the film -  are presented.
In many interesting cases, such as in FCC (001) films and in FCC(110) films,
the energies of most electronic
states in the film can be analytically obtained from the corresponding
energy band structure of the bulk: For each energy band of the bulk
and each two-dimensional wavevector 
$\hat{\bf k}$ in the film plane, there are $N_3-1$ electronic states
in the film whose energy is dependent on the film thickness $N_3$ but 
not on the film boundary $\tau_3$
and maps the energy band of the bulk exactly and one electronic state whose
energy is dependent on $\tau_3$ but not on $N_3$ and is above or at the
highest energy in that energy band with that $\hat{\bf k}$.
This approach can be further extended to obtain exact and general
results on electronic states in quantum wires and quantum dots.
\newpage
\par
Bloch theorem has been playing a central role in
our current understanding on the electronic structures of
crystals in modern solid state physics.
However, any real crystal always has a finite size and does not have 
the translational invariance on which Bloch theorem is based. 
A clear understanding of the properties
of electronic states in real crystals of finite size has 
both theoretical and practical significant importance. 
Nevertheless, the lack of translational invariance has been 
a major obstacle in obtaining exact and general results on the
electronic states in crystals of finite size.
Thus most previous theoretical investigations on the
electronic structure of finite crystals were based on 
approximate and/or numerical approaches and were usually on a specific 
material and/or based on a specific model\cite{rvs}.
Recently the author obtained exact results on 
the electronic states in ideal general one dimensional crystals bounded at
$\tau$ and $\tau+L$ - where $\tau$ is a real number and $L$ is a length
equal to a positive integer times of the potential period $a$ - by using 
a differential equation theory approach\cite{syr1,syr0,eas}. 
One of the major results of \cite{syr1} is that the electronic states in 
one dimensional crystals of finite length can be classified as
$L$- dependent or $\tau$-dependent - the energy of the electronic
state is dependent on either the length of the crystal $L$ or
the crystal boundary $\tau$. It also shows that the
obstacle due to the lack of translational invariance in one dimensional
finite crystals in fact can be circumvented. 
The basis of \cite{syr1} is a clear understanding of the $zeros$ of solutions
of $ordinary$ differential equations with periodic 
coefficients obtained by mathematicians\cite{eas}. 
Now we 
extend our investigations into the three dimensional cases. In this work, we 
treat the simplest case - the electronic states in ideal free standing films. 
\par 
The most significant difference between the problem treated 
in this work and \cite{syr1} is, 
the corresponding Schr$\ddot{\rm o}$dinger equation for the electronic
states in films is a $partial$ differential equation and thus the problem
is now a more difficult one. Furthermore, 
for the zeros of solutions of three dimensional $partial$ differential 
equation with periodic coefficients, much less was understood.
Nevertheless, based on the results of a new eigenvalue problem - an extension
of a mathematical theorem in \cite{eas} - 
we demonstrate that in many interesting cases 
exact and general results on how the eigenvalues of 
electronic states in ideal free standing films
depend on the thickness can be analytically obtained, many eigenvalues
can be directly obtained from the
energy band structure of the bulk. Again, the major obstacle 
due to the lack of translational invariance can be circumvented.
We will mainly be interested in films with face-center cubic(FCC)
lattices, including diamond structure and zinc-blend structure. 
\par
The Schr$\ddot{\rm o}$dinger equation for
three dimensional crystals can be written as 
\begin{equation}
-\nabla^2 y({\bf x}) + [ v({\bf x}) - \lambda ] y({\bf x}) = 0,
~~~~~~
\end{equation}
where 
$$
v({\bf x} + {\bf a}_1) = v( {\bf x}),~~~~~v({\bf x} + {\bf a}_2) = v( {\bf x}),
~~~~~v({\bf x} + {\bf a}_3) = v( {\bf x}).
$$
${\bf a_1}$, ${\bf a_2}$ and ${\bf a_3}$ are three
primitive vectors of the 
crystal. The corresponding primitive vectors in {\bf k} space
are denoted as ${\bf b_1}$, ${\bf b_2}$ and ${\bf b_3}$ and  
${\bf a}_i \cdot {\bf b}_j = \delta_{i,j}$.
The position vector ${\bf x}$ can be written as 
$ {\bf x} = x_1 {\bf a_1} + x_2 {\bf a_2} + x_3 {\bf a_3}$ 
and the ${\bf k}$ vector as
$ {\bf k} = k_1 {\bf b_1} + k_2 {\bf b_2} + k_3 {\bf b_3}$.
\par
The eigenfunctions of (1) satisfying the condition
\begin{equation}
\phi({\bf k}, {\bf x} + {\bf a}_i ) = 
\phi({\bf k}, {\bf x}) exp(i k_i)  ~~~~  -\pi < k_i \leq \pi  ~~~ i = 1, 2, 3
\end{equation}
are three dimensional Bloch functions.
In this work the three dimensional Bloch functions and the 
energy bands are denoted as
$ \phi_n({\bf k},{\bf x}) $ and $ \lambda_n( {\bf k})$:
$
\lambda_0({\bf k}) \leq \lambda_1({\bf k}) \leq \lambda_2({\bf k}) \leq .....
$
The energy band structure in the Cartesian system is denoted as
$\lambda_n(k_x, k_y, k_z)$.
\par
For the electronic states in ideal free standing films we assume that
the atomic positions in the film are the same as in the
bulk, the potential $v({\bf x})$ $inside$ the film is the same as in (1)
and all the electronic states are confined in the film.
We choose primitive vectors ${\bf a}_1$ and ${\bf a}_2$ in the film plane
and use $\hat{\bf k}$ to express the two dimensional wavevector:
$\hat{\bf k} = \hat{k}_1 \hat{\bf b}_1 + \hat{k}_2 \hat{\bf b}_2$. 
$\hat{\bf b}_1 $ and $ \hat{\bf b}_2$ are in the film plane and
${\bf a}_i \cdot \hat {\bf b}_j = \delta_{i,j}$.
\par
Suppose $A$ is a parallelogram which has ${\bf a}_i$ forming the sides
which meet at a corner and has the bottom defined by $\tau_3{\bf a}_3$\cite{t3}.
The function set $\hat{\phi}(\hat{\bf k}, {\bf x}; \tau_3)$ is defined by 
the condition
$$
\hat{\phi}(\hat{\bf k}, {\bf x} + {\bf a}_i; \tau_3 ) 
= \hat{\phi} (\hat{\bf k}, {\bf x}; \tau_3) exp(i k_i) ~~~~  
-\pi < k_i \leq \pi  ~~~ i = 1, 2 
$$
\begin{equation}
\hat{\phi} (\hat{\bf k}, {\bf x}; \tau_3 ) = 0
 ~~~~~~~~~~~~~~~~~~~~~~~~~~~~~~~if~ {\bf x}~ \in~ \partial A_3
\end{equation}
here $\partial A_i$ means two opposite faces of $\partial A$ determined by 
the beginning and the end of ${\bf a}_i$.
The eigenvalues and eigenfunctions of (1) with the condition (3) are denoted by 
$\hat{\lambda}_n(\hat{\bf k}; \tau_3)$ and 
$\hat{\phi}_n(\hat{\bf k}, {\bf x}; \tau_3)$,
where $\hat{\bf k}$ is the two dimensional wavevector and $n~=~0,~1,...$.
\par
For the eigenvalues of two different eigenvalue problems defined by (2) and (3),
we have the following theorem(See the Appendix):
\begin{equation}
\hat{\lambda}_n(\hat{\bf k}; \tau_3) \geq \lambda_n({\bf k}),
~~~~~~~~~~~~ for~({\bf k} - \hat{\bf k}) \cdot {\bf a}_i = 0, ~~i =1, 2.
\end{equation}
Eq. (4) is more or less like the Theorem 3.1.2 in \cite{eas} in 
the one dimensional case, but not as strong as the latter: No upper limit
of $\hat{\lambda}_n(\hat{\bf k}; \tau_3)$ is given except
$\hat{\lambda}_n(\hat{\bf k}; \tau_3) \leq
\hat{\lambda}_{n+1}(\hat{\bf k}; \tau_3)$.
\par
Note in Eqs. (3) and (4), ${\bf k}$ is a three dimensional wavevector and
$\hat{\bf k}$ is a two dimensional wavevector. In Eq. (4) ${\bf k}$ and 
$\hat{\bf k}$ have the same two dimensional wavevector in the film plane.
For each energy band $n$ and each $\hat{\bf k}$, there is 
one $\hat{\phi}_n(\hat{\bf k}, {\bf x}; \tau_3)$.
Eq. (4) is true for $any$ $\tau_3$, thus it indicates that in general a 
Bloch function 
$\phi_n({\bf k},{\bf x})$ does not have a "zero surface" $\tau_3(x_1,x_2)$, 
except when it happens to be one of 
$\hat{\phi}_{n'}(\hat{\bf k}, {\bf x}; \tau_3)$.
\par
For the electronic states in a film with $N_3$ layers in the ${\bf a}_3$ 
direction, we look for the the eigenvalues 
$\hat{\Lambda} $ and eigenfunctions
$\hat{\psi}( \hat{\bf k}, {\bf x})$ of the following two equations:
\begin{equation}
 - \nabla^2 \hat{\psi}(\hat{\bf k}, {\bf x}) 
+ [ v({\bf x} ) - \hat{\Lambda} ] 
\hat{\psi}(\hat{\bf k}, {\bf x}) = 0,
~~~~~~~~~~~~~~~~~~~~~~~~if~~\tau_3~<~x_3~<~\tau_3 + N_3
\end{equation}
and
\begin{equation}
\hat{\psi}(\hat{\bf k}, {\bf x}) = 0,~~~~~~~~~~~~~~~~~~~~~~~~~~~~~~~~~~~~~~
 ~~~~if~~x_3~\leq~\tau_3~~or~~ x_3 ~\geq~\tau_3 +N_3~~
\end{equation}
where 
$ x_3 =\tau_3 $ indicating the bottom of the
film  and $ N_3~$ is a positive integer
indicating the film thickness. 
These electronic states $\hat{\psi}(\hat{\bf k}, {\bf x})$ 
are two dimensional Bloch waves in the film plane with 
additional index(es) indicating the confinement in the third direction. 
\par
One type of non-trivial solutions of (5) and (6) can be obtained from (3) 
by assigning
$$
\hat{\psi}_n( \hat{\bf k}, {\bf x}; \tau_3) 
=c_{N_3} \hat{\phi}_n (\hat{\bf k}, {\bf x}; \tau_3),
~~~~~~~~~~~~~~~~~if~~\tau_3~<~x_3~<~\tau_3 + N_3
$$
\begin{equation}
~~~~~~~~~~~~~~~~~~~~~~ = 0,~~~~~~~~~~~~~
 ~~~~~~~~~~~~~~~if~~x_3~\leq~\tau_3~~or~~ x_3~\geq~\tau_3 +N_3~~
\end{equation}
where $c_{N_3}$ is a normalization constant.
The corresponding eigenvalue 
$\hat{\Lambda}_n(\hat{\bf k}; \tau_3) =
\hat{\lambda}_n(\hat{\bf k} ; \tau_3) $
is dependent on $\tau_3$ but not on $N_3$.
A consequence of (4) is that
for each energy band index $n$ and each two dimensional wavevector
$\hat{\bf k}$, there is one such solution (7) of Eqs. (5) and (6).
We will further discuss this types of solutions later.
Now we try to find out $other$ solutions of (5) and (6).
\par
The problem is simpler if the band structure of the crystal has the 
following symmetry
\begin{equation}
\lambda_n(k_1 {\bf b}_1 + k_2 {\bf b}_2 + k_3 {\bf b}_3) =
\lambda_n( k_1 {\bf b}_1 + k_2 {\bf b}_2 - k_3 {\bf b}_3).
\end{equation}
If this is the case, we note that
$$
f_{n,k_1,k_2,k_3} ( {\bf x}; \tau_3 ) = c_{n,k_1,k_2,k_3} 
\phi_n(k_1{\bf b}_1 + k_2{\bf b}_2 + k_3{\bf b}_3, {\bf x} ) 
+ c_{n,k_1,k_2,-k_3} 
\phi_n(k_1{\bf b}_1 + k_2{\bf b}_2 - k_3{\bf b}_3, {\bf x} ),
~~~~~~~~~~~~~ 
$$
where $c_{n,k_1,k_2,k_3}$ and $c_{n,k_1,k_2,-k_3}$ are 
constant coefficients and are not 
zero, is a non trivial solution of (5) due to (8).
Non-trivial solutions of (5) and (6) can be obtained by assuming
$$
\hat{\psi}_{n,j_3}( \hat{\bf k}, {\bf x}; \tau_3) 
=f_{n,k_1,k_2,\kappa_3} ( {\bf x}; \tau_3 ) 
~~~~~~~~~~~~~~~~~~~~if~~\tau_3~<~x_3~<~\tau_3 + N_3
$$
\begin{equation}
~~~~~~~~~~~~~~~~~~~~~~~~~~ = 0,~~~~~~~~~~~~~
 ~~~~~~~~~~~~~~~if~~x_3~\leq~\tau_3~~or~~ x_3~\geq~\tau_3 +N_3~~
\end{equation}
where $\hat{\bf k} = k_1 \hat{\bf b}_1 + k_2 \hat{\bf b}_2$, 
or $\hat{k}_1 = k_1$ and $\hat{k}_2 = k_2$, and 
\begin{equation}
\kappa_3 = j_3~\pi/N_3,~~~~~~~~~j_3~=~1,~2,~3,......N_3-1.
\end{equation}
Here $j_3$ can be considered as a sub-band index. Because from (10) we have 
$ e^{i \kappa_3 N_3} - e^{- i \kappa_3 N_3} = 0$,
then we can always choose $c_{n,k_1,k_2,\kappa_3}$ 
and $c_{n,k_1,k_2,-\kappa_3}$ which are not zero to make\cite{t3}
$
f_{n,k_1,k_2,\kappa_3} (x_1 {\bf a_1} + x_2 {\bf a_2} + \tau_3 {\bf a_3})
 = f_{n,k_1,k_2,\kappa_3} 
( x_1 {\bf a_1} + x_2 {\bf a_2} + (\tau_3 + N_3) {\bf a_3} ) = 0.
$
Therefore $\hat{\psi}_{n,j_3}(\hat{\bf k}, {\bf x}; \tau_3)$ 
defined in (9)
are continuous functions satisfying (5) and (6) with eigenvalues given by
\begin{equation}
\hat{\Lambda}_{n,j_3}( k_1 \hat{\bf b}_1 + k_2 \hat{\bf b}_2) = 
\lambda_n( k_1 {\bf b}_1 + k_2 {\bf b}_2 + \kappa_3 {\bf b}_3) 
\end{equation}
Each eigenvalue $\hat{\Lambda}_{n,j_3} (\hat{\bf k})$   
for this case, is a function of $N_3$, the film thickness.
But they all do not depend on the film boundary $\tau_3$. 
The eigenvalues $\hat{\Lambda}_{n, j_3} (\hat{\bf k})$ of the electronic  
states in the film (11) map the  band structure of the bulk
$ \lambda_n( k_1 {\bf b}_1 + k_2 {\bf b}_2 + k_3 {\bf b}_3)$ 
exactly.
\par
However, for many crystals such as important semiconductors with diamond
structure or zinc-blend structure, in general 
(8) is not true. Thus we do not have (11) for general 
$k_1$ and $k_2$. We have to treat each film case separately. In the
following we treat FCC (001) films and FCC (110) films as two examples.\par
\par
We discuss the FCC (001) films first. 
The primitive vectors can be chosen as 
${\bf a}_1 = a/2~(1, -1, 0)$ and ${\bf a}_2 = a/2~(1, 1, 0)$,
${\bf a}_3 = a/2~(1, 0, 1)$ and thus
${\bf b}_1 = 1/a~(1, -1, -1)$,
${\bf b}_2 = 1/a~(1, 1, -1)$,
and ${\bf b}_3 = 1/a~(0, 0, 2)$. 
Three Cartesian components of the wavevector ${\bf k}$ can be obtained as
$k_x = (k_1 + k_2)/a $, $k_y = (- k_1 + k_2)/a$ and 
$k_z = (- k_1 - k_2 + 2 k_3 )/a$.
\par
In general, we do not have (8) for FCC crystals.
Nevertheless, due to the high symmetry of the FCC band structures,
we are still able to obtain an analytical expressions for the electronic
states in FCC (001) films, similar to (11).\par
In fact, we can have many different ways to choose the primitive vectors.
For example, we can choose a new primitive vector system as
 ${\bf a}'_1 = {\bf a}_1 $,
 ${\bf a}'_2 = {\bf a}_2 $,
 ${\bf a}'_3 = m_1 {\bf a}_1 + m_2 {\bf a}_2 + {\bf a}_3$,
here $m_1$ and $m_2$ are integers.
The new primitive vectors in ${\bf k}$ space are 
 ${\bf b}'_1 = {\bf b}_1 - m_1 {\bf b}_3$,
 ${\bf b}'_2 = {\bf b}_2 - m_2 {\bf b}_3$,
 ${\bf b}'_3 = {\bf b}_3 $.
In principle, the primitive vector systems given by ${\bf a}_i$ and
${\bf a}'_i$ are equivalent.
The wavevector ${\bf k}$ can be expressed as 
either $ k_1{\bf b}_1 + k_2{\bf b}_2 + k_3{\bf b}_3 $ or
$ k'_1{\bf b}'_1 + k'_2{\bf b}'_2 + k'_3{\bf b}'_3 $
and
$k_1 = k'_1 $,
$k_2 = k'_2 $,
$k_3 = - m_1 k'_1 - m_2 k'_2 + k'_3$.
Thus for a specific wavevector ${\bf k}$, in two different primitive vector 
systems used here, both $k_1$ and $k_2$ are unchanged.\par
In FCC (001) films we have
$\hat{\bf b}_1 = 1/a (1, -1, 0) $ and $\hat{\bf b}_2 = 1/a (1, 1, 0)$,
$ \hat{k}_1 = k_1$ and $\hat{k}_2 = k_2$. 
Our purpose is to find $\hat{\Lambda}_{n,j_3} (\hat{\bf k})$ for 
different $\hat{\bf k}$.\par 
For FCC crystals we have
$\lambda_n(k_x, k_y, k_z) = \lambda_n(k_x, k_y, -k_z) $
for any $k_x$ and $k_y$. We have also that $k_x = (k_1 + k_2)/a $, 
$k_y = (- k_1 + k_2)/a$, $k_z = (- k_1 - k_2 + 2 k_3)/a = 
( - (2 m_1 +1 ) k_1 - (2 m_2 +1) k_2 + 2 k'_3)/a$.
If for two specific $k_1$ and $k_2$, we can find two $m_1$ and $m_2$,
to make 
\begin{equation}
(2 m_1 +1 ) k_1 + (2 m_2 +1) k_2 = 0,
\end{equation}
then in the primitive vector system specifying by 
 ${\bf a}'_3 = m_1 {\bf a}_1 + m_2 {\bf a}_2 + {\bf a}_3$,
we have
$$
\lambda_n( k_1 {\bf b}'_1 + k_2 {\bf b}'_2 + k'_3 {\bf b}'_3) =
\lambda_n( k_1 {\bf b}'_1 + k_2 {\bf b}'_2 - k'_3 {\bf b}'_3) 
$$
and thus
$$
\hat{\Lambda}_{n,j_3} (k_1 \hat{\bf b}_1 + k_2 \hat{\bf b}_2)  
= \lambda_n ( k_1 {\bf b}'_1 + k_2 {\bf b}'_2 + \kappa_3 {\bf b}'_3).
$$
This in fact is\cite{do1}
\begin{equation}
\hat{\Lambda}_{n,j_3} (k_1 \hat{\bf b}_1 + k_2 \hat{\bf b}_2) 
=\lambda_n ( k_1 \hat{\bf b}_1 +  k_2 \hat{\bf b}_2 + \kappa_3 {\bf b}_3),
\end{equation}
where $\kappa_3 = j \pi/N_3,~j_3~=~1,~2,...N-1$.
Note (13) does not depend on $m_1$ and $m_2$ any more.\par
There are some pairs of $k_1$ and $k_2$ for which
the condition (12) can not be true. Nevertheless,
in a small circle centered in an any specific pair $k_1$ and $k_2$, there are 
always infinite number of pairs $k_{1,c}$ and $k_{2,c}$, 
which can be as close to $k_1$ and $k_2$ as needed, 
for each pair we can find two integers $m_1$ and
$m_2$, to make $(2m_1 +1) k_{1,c} + (2 m_2+1) k_{2,c} = 0$. 
Thus (13) will be true for each such pair $k_{1,c}$ and $k_{2,c}$. Because both
$ \hat{\Lambda}_{n,j_3} (k_1 \hat{\bf b}_1 + k_2 \hat{\bf b}_2) $ 
and $\lambda_n ( k_1 \hat{\bf b}_1 +  k_2 \hat{\bf b}_2 + \kappa_3 {\bf b}_3)$
are continuous functions of $k_1$ and $k_2$, therefore (13) must be true for 
any $k_1$ and $k_2$. The corresponding eigenfunction can be written as
$\hat{\psi}_{n,j_3} (\hat{\bf k}, {\bf x}; \tau_3)$.\par
In the Cartesian system, Eq. (13) for (001) films can be written as: 
\begin{equation}
\hat{\Lambda}_{n,j_3} ( k_x, k_y) = 
\lambda_n (k_x, k_y, 2 \kappa_3 ) 
\end{equation}
for any $k_x$ and $k_y$, where $\kappa_3 = j_3 \pi/N_3 $.\par 
Eqs. (13) and (14) can be used for Si (001) films. 
Zhang and Zunger\cite{zz} calculated the electronic structure in
thin Si (001) films using a pseudopotential method. Their results for the
even number $N_f$ of monolayer can be directly compared with our general
analytical results: Our $N_3$ is equal to their $N_f/2$. 
Their "central observation" is that 
the energy spectrum of electronic states in Si quantum films
($N_f = 12$) maps the energy band structure of Si approximately.
The equation (9) in \cite{zz}, which Zhang and Zunger obtained from 
their numerical results, is a special case 
of Eq. (14) when $k_x = k_y = 0$. Therefore Eq. (14) is a more general result
obtained analytically.
\par
The FCC (110) films can be very similarly discussed. 
\par
The primitive vectors can
be chosen as ${\bf a}_1 = a~( 0, 0, 1)$ and ${\bf a}_2 = a/2~ (1, -1, 0)$,
${\bf a}_3 = a/2~(0, 1, 1)$.  Correspondingly we have
${\bf b}_1 = 1/a~(- 1, - 1, 1)$,
${\bf b}_2 =  1/a~ ( 2, 0, 0) $,
${\bf b}_3 = 1/a~(2, 2, 0)$, 
and $\hat{\bf b}_1 = 1/a (0, 0, 1) $, $\hat{\bf b}_2 = 1/a (1, -1, 0)$.
By using very similar argument as we obtain Eq. (13) for FCC (001) films, but 
using that for FCC crystals $\lambda_n(k_3 + k_2, k_3 - k_2, k_1) 
= \lambda_n(-k_3 + k_2, -k_3 - k_2, k_1) $ is true
for any $k_1$ and $k_2$, we can obtain that for the electronic states
in the $N_3$ layers of FCC (110) films, 
$\hat{\Lambda}_{n,j_3} ( k_1 \hat{\bf b}_1 + k_2 \hat{\bf b}_2) 
=\lambda_n ( k_1 \hat{\bf b}_1 + k_2 \hat{\bf b}_2 + \kappa_3 {\bf b}_3) $
for any $k_1$ and $k_2$.  Here 
$\kappa_3 = j_3 \pi/N_3 $, $j_3$ is the sub-band index: $j_3 = 1, 2,...N_3-1$. 
This is exactly the same as Eq. (13). Thus Eq. (13) is correct for both 
FCC (001) films and FCC (110) films. The difference 
lies on that $\hat{\bf b}_1, \hat{\bf b}_2$ and 
${\bf b}_3$ are different in two cases. 
For FCC (110) films, Eq. (13) can also be written as
\begin{equation}
\hat{\Lambda}_{n,j_3} ( k_1 \hat{\bf b}_1 + k_2 \hat{\bf b}_2) 
= \lambda_n ( (2 \kappa_3 - k_2)/a, (2 \kappa_3 + k_2)/a, k_1/a ) 
\end{equation}
In the cases where $k_1 = k_2 = 0$, we have $ \hat{\Lambda}_{n,j_3} (0) = 
\lambda_n (j_3/N_3~2 \pi/a, j_3/N_3~2 \pi/a, 0 ) $.
This is what Zhang and Zunger\cite{zz} obtained from 
their numerical results on Si (110) films
due to that our $N_3$ is equal to their $ N_f $ in (110) films.
\par
Much previous work also indicates that the eigenvalues of Bloch 
states in films map $closely$ the dispersion relations of the unconfined Bloch 
waves\cite{pop}. Now we understand that the mapping in fact is $general$ 
and $exact$ in the cases treated here.
\par
Franceschetti and Zunger\cite{fz} investigated the $\Gamma - X$ transition in
GaAs free standing (110) films, quantum wires and quantum dots. 
According to Eq. (15) we have analytically that for (110) films
$$
E^{\Gamma}_{film}=
\frac{\hbar^2}{2m} \lambda_c(1/N_3~2 \pi/a, 1/N_3~2 \pi/a, 0),
~~
E^{X}_{film}=
\frac{\hbar^2}{2m} \lambda_c (1/N_3~2 \pi/a, 1/N_3~2 \pi/a, k_{min}),
$$
where $c$ denotes the lowest conduction band, $k_{min}$ is the location 
of the "X" valley on the $k_z$ axis and $N_3$ is the number of mono-layers.
\par
Each $\hat{\psi}_n(\hat{\bf k}, {\bf x}; \tau_3)$ defined in (7) is an
electronic state in the film whose energy 
$\hat{\Lambda}_{n}( \hat{\bf k}; \tau_3) $ is dependent on the film boundary
$\tau_3$ but independent of the thickness of the film $N_3$. 
In one-dimensional crystals of finite length\cite{syr1}
the energy bands $\lambda_n(k)$ do not overlap and a 
$\tau$-dependent eigenvalue is either inside the band gap or at the band edge.
In films the energy bands $\lambda_n({\bf k})$ usually overlap and
$\hat{\Lambda}_{n}( \hat{\bf k}; \tau_3) $ 
is either above or at the highest energy of the Bloch function
with that $n$ and that $\hat{\bf k}$.\par
If in Eq. (4) $\hat{\lambda}_n(\hat{\bf k}; \tau_3) = 
\lambda_n({\bf k})$, 
$\hat{\Lambda}_{n}( \hat{\bf k}; \tau_3)$ is $equal~to$ the
highest energy for that energy band and that $\hat{\bf k}$.
It means that the corresponding eigenfunction of (3)
$\hat{\phi}_n(\hat{\bf k}, {\bf x})$ is a Bloch function
$\phi_n({\bf k}, {\bf x})$.
Examples of this case were observed, for example,
in numerical results in the Fig. 3 and Fig. 4 of \cite{zz}, such as the
one of the double degenerated $X_1$ state in the valence band, the
constant energy confined band edge state at the top of valence band, etc.
Other examples are the constant energy confined band edge states
at the top of the valence band observed in numerical calculations
for Si (110) films\cite{zz} and GaAs (110) film\cite{fz} etc.
This case happens when the corresponding Bloch function
$\phi_n({\bf k}, {\bf x})$ with $\hat{\bf k}$ has a "zero surface" at 
$x_3 = \tau_3$ and thus has "zero surfaces" at $x_3 = \tau_3 + j $, 
where $ j = 1, 2,...... N_3$.
\par
In the cases where
$
\hat{\lambda}_n(\hat{\bf k}; \tau_3) > \lambda_n({\bf k}),
$
none of the Bloch function $\phi_n({\bf k}, {\bf x})$ with that $n$ and that
$\hat{\bf k}$ has a "zero surface" at $x_3 = \tau_3$.
Then 
$\hat{\psi}_n(\hat{\bf k}, {\bf x}; \tau_3)$ could be a surface state.
In ideal one dimensional crystals of finite length we have proved that 
there is $at~most~ one$ surface state in each band gap\cite{syr1}.
Eq. (4) shows, there
is $at~most~one$ surface energy band $\hat{\Lambda}_n(\hat{\bf k}, \tau_3)$
for each bulk energy band $\lambda_n({\bf k})$, even a film always has $two$
surfaces. Nevertheless, due to that in three dimensional crystals below the 
major band gap there are several energy bands
overlapping, such as the energy bands $n=1,2,3$ in Si or GaAs etc, thus
in the major band gap there could be more than one surface energy bands.
However, for the band gap $inside~ the~valence~band$ of III-V and II-VI
semiconductors between the band $n=0$ and the band $n=1$, we predict that there
is $at~most~one$ surface energy band $\hat{\Lambda}_0(\hat{\bf k}; \tau_3)$ 
for a $perfectly~even$ (001) or (110) film, as a direct consequence of Eq. (4). 
\par
In summary, we have demonstrated that for interesting free standing films,
such as ideal FCC (001) films and FCC (110) films, exact and general results
on the eigenvalues of the electronic states in the film can be obtained:
For the film bounded between $\tau_3 {\bf a}_3$
and $ (N_3 + \tau_3) {\bf a}_3$,
there are $N_3 -1$ electronic states 
$\hat{\psi}_{n, j_3} (\hat{\bf k}, {\bf x}; \tau_3)$ 
for each energy band $n$ and each
two dimensional wavevector $\hat{\bf k}$, whose energy 
$\hat{\Lambda}_{n, j_3}(\hat{\bf k})$ is dependent 
on the thickness of the film $N_3$ but not on the film boundary $\tau_3$. 
The eigenvalues of these electronic states in the film map the energy bands
$\lambda_n({\bf k})$ exactly;
There is one electronic state 
$\hat{\psi}_n(\hat{\bf k}, {\bf x}; \tau_3)$ in the film for each
$n$ and $\hat{\bf k}$ whose energy $\hat{\Lambda}_n(\hat{\bf k}; \tau_3)$
is dependent on the film boundary
$\tau_3$ but not on the film thickness $N_3$. 
\par
This approach can be further extended to obtain exact and general
results on electronic states in
quantum wires and quantum dots\cite{syr2}. For example, it is found that
for a quantum dot which are formed by FCC (001) or (110) films
further truncated in two more directions and thus bounded between 
$\tau_1 {\bf a}_1$ and $(\tau_1 + N_1) {\bf a}_1$ and
$\tau_2 {\bf a}_2$ and $(\tau_2 + N_2) {\bf a}_2$ and
$\tau_3 {\bf a}_3$ and $(\tau_3 + N_3) {\bf a}_3$, 
for each energy band there are $(N_1-1)(N_2-1)(N_3-1)$ bulk states,
$(N_1-1)(N_2-1) + (N_2-1)(N_3-1) + (N_3-1)(N_1-1)$ surface states,
$(N_1-1) + (N_2-1) + (N_3-1)$ "side states" and one "conner state" in spite
of that the crystal has six faces, twelve sides and eight corners.

\newpage
\begin{center}
Appendix
\end{center}
The Eq. (4) in the text is essentially an extension of the theorem (6.3.1)
in \cite{eas} and its proof is similar, with some differences.\par
We choose $\phi_n( {\bf k}, {\bf x})$
to be normalized over $A$: 
$
\int_{A} \phi_n( {\bf k}; {\bf x})~~\phi^*_n( {\bf k}; {\bf x}) d{\bf x} = 1.
$
\par
We denote $F$ as the set of all complex-valued functions $f({\bf x})$ which are
continuous in $A$ and have piecewise continuous first-order partial derivatives
in $A$. The Dirichlet integral $J(f,g)$ in three dimensions is
defined by
$$
J(f, g) = \int_A \{ \nabla f({\bf x}) \cdot \nabla g^*({\bf x}) 
+ v ({\bf x}) f({\bf x}) g^*({\bf x}) \} d {\bf x}
\eqno (A.1)
$$
for $f({\bf x})$ and $g({\bf x})$ in $F$.
If in $(A.1)$ $g({\bf x})$ also has piecewise continuous
second-order partial derivatives in $A$, from Green's theorem we have
$$
J(f, g) =  \int_A f({\bf x}) \{ - \nabla^2 g^*({\bf x}) 
+ v ({\bf x}) g^*({\bf x}) \} d {\bf x}
+ \int_{\partial A} f \frac{\partial g^*}{\partial n} d S,
\eqno (A.2)
$$
where $\partial A$ denotes the boundary of $A$, $\partial/\partial n$ denotes
derivative along the outward normal to $\partial A$, and $d S$ denotes an
element of surface area of $\partial A$.
\par
If $f({\bf x})$ and $g({\bf x})$ satisfy the conditions Eq. (2) in the text, 
the integral over $\partial A$ is zero because the integrals over
opposite faces of $\partial A$ cancel out. In particular, when
$g({\bf x}) = \phi_n ({\bf k}, {\bf x}) $, $(A.2)$ gives
$$
J(f, g) =  \lambda_n({\bf k}) 
\int_A f({\bf x}) \phi^*_n({\bf k}, {\bf x}) d{\bf x}.
$$
Thus $ J(\phi_m({\bf k},{\bf x}), \phi_n({\bf k},{\bf x})) 
  = \lambda_n({\bf k})$ if $ m = n$
and
$ J(\phi_m({\bf k},{\bf x}), \phi_n({\bf k},{\bf x})) = 0$ if $ m \neq n$.
\par
Now we consider the function set 
$\hat{\phi}(\hat{\bf k}, {\bf x}; \tau_3)$ which satisfy 
the conditions (3) in the text. We also choose 
$\hat{\phi}(\hat{\bf k}, {\bf x}; \tau_3)$ to be normalized over $A$:
$ \int_A \hat{\phi}(\hat{\bf k}, {\bf x}; \tau_3)
\hat{\phi}^*(\hat{\bf k}, {\bf x}; \tau_3) d {\bf x} = 1$. 
\par
Note if $ f({\bf x}) = \hat{\phi}(\hat{\bf k}, {\bf x}; \tau_3)$, 
and $g({\bf x}) = \phi_n ({\bf k}, {\bf x}) $, the integral over $\partial A$
in $(A.2)$ is also zero because the integral over two opposite faces 
of $\partial A_1$ and $\partial A_2$ cancel out due to that
$({\bf k} - \hat{\bf k}) \cdot {\bf a}_i = 0 $ for $i~=~1,~2$
and the integral over 
each face of $\partial A_3$ is zero, due to that $f({\bf x}) = 0$
when ${\bf x} \in \partial A_3$.
\par
Thus
$$
J(\hat{\phi}(\hat{\bf k}, {\bf x}; \tau_3), \phi_n( {\bf k}, {\bf x})) =
\lambda_n({\bf k}) f_{n} ({\bf k})
$$
where
$$
f_{n}({\bf k}) = \int_A 
\hat{\phi}(\hat{\bf k}, {\bf x}; \tau_3) 
\phi^*_n( {\bf k}, {\bf x}) d{\bf x},
$$
and 
$$
\sum_{n=0}^{\infty} | f_{n} ({\bf k})|^2  = 1.
$$
An important property of the function 
$\hat{\phi}(\hat{\bf k}, {\bf x}; \tau_3) $ defined by (3) in the text
is
$$
J(\hat{\phi}(\hat{\bf k}, {\bf x}; \tau_3), 
\hat{\phi}(\hat{\bf k}, {\bf x}; \tau_3) )
\geq
\sum_{n=0}^{\infty} \lambda_n({\bf k}) | f_{n} ({\bf k})|^2 .
\eqno (A.3)
$$
To prove $(A.3)$, we assume $v({\bf x}) \geq 0$ first.
Then $J(f,f) \geq 0$ from $(A.1)$ for any $f$ in $F$. 
Thus for any positive integer $N$ we have
$$
J(\hat{\phi}(\hat{\bf k}; {\bf x}; \tau_3) - \sum_{n=0}^{N} 
f_{n} ({\bf k}) \phi_n( {\bf k}, {\bf x}),
~\hat{\phi}(\hat{\bf k}, {\bf x}; \tau_3) - \sum_{n=0}^{N} 
f_{n} ({\bf k}) \phi_n( {\bf k}, {\bf x}))
 \geq 0.
$$
That is 
$$
J(\hat{\phi}(\hat{\bf k}, {\bf x}; \tau_3), 
\hat{\phi}(\hat{\bf k}; {\bf x}; \tau_3) )
\geq
\sum_{n=0}^{N} \lambda_n({\bf k}) | f_{n} ({\bf k})|^2. 
$$
N can be as large as it needs, therefore
$$
J(\hat{\phi}(\hat{\bf k}, {\bf x}; \tau_3), 
\hat{\phi}(\hat{\bf k}, {\bf x}; \tau_3)) 
\geq
\sum_{n=0}^{\infty} \lambda_n({\bf k}) | f_{n} ({\bf k})|^2 .
~~~~~~~~~~~~~~ if~ v({\bf x}) \geq 0
\eqno (A.4)
$$
To prove $(A.3)$ without the assumption that $v({\bf x}) \geq 0$, let $v_0$
be a constant which is sufficient large to make 
$ v({\bf x}) + v_0 \geq 0$ in $A$. Then Eq. (1) in the text
can be written as
$$ 
- \nabla^2 y({\bf x}) + [ V({\bf x}) - \Lambda ] y({\bf x}) = 0
\eqno (A.5)
$$
where
$ V({\bf x}) = v({\bf x}) + v_0 $ and $\Lambda = \lambda + v_0$. 
Since in $(A.5)$ $V({\bf x}) \geq 0$, due to $(A.4)$ we have
$$
\int_A \{ \nabla \hat{\phi}(\hat{\bf k}; {\bf x} ; \tau_3) \cdot 
\nabla \hat{\phi}^*(\hat{\bf k}; {\bf x} ; \tau_3)  
+ [v ({\bf x}) + v_0] 
\hat{\phi}(\hat{\bf k}; {\bf x} ; \tau_3)  
\hat{\phi}^*(\hat{\bf k}; {\bf x} ; \tau_3)  
\} d {\bf x}
~~~~~~~~~~~~~~~~~~~~~~~
$$
$$
~~~~~~~~~~~~~~~~~~~~~~~~~~~~~~~~~~~~~~~~~~~~~~~~~~~~~~~~~~~~~~~~~~~~~~~~~~~
\geq
\sum_{n=0}^{\infty} (\lambda_n({\bf k}) + v_0) | f_{n} ({\bf k})|^2 
$$
That is
$$
\int_A \{ \nabla \hat{\phi}(\hat{\bf k}; {\bf x} ; \tau_3) \cdot 
\nabla \hat{\phi}^{*}(\hat{\bf k}; {\bf x} ; \tau_3)  
+ v ({\bf x}) 
\hat{\phi}(\hat{\bf k}; {\bf x} ; \tau_3)  
\hat{\phi}^{~*}(\hat{\bf k}; {\bf x} ; \tau_3)  
\} d {\bf x}
\geq
\sum_{n=0}^{\infty} \lambda_n({\bf k}) | f_{n} ({\bf k})|^2 
$$
This is $(A.3)$. On the basis of $(A.3)$ we can prove Eq. (4) in the text.
\par
We consider
$$
\hat{\phi}(\hat{\bf k}, {\bf x}; \tau_3) = 
c_0 \hat{\phi}_0 (\hat{\bf k}, {\bf x}; \tau_3 ) +
c_1 \hat{\phi}_1 (\hat{\bf k}, {\bf x}; \tau_3 ) 
+ ....
+ c_n \hat{\phi}_n (\hat{\bf k}, {\bf x}; \tau_3 ) 
$$
and choose $n+1$ constants $c_i$ to make
$$
\sum_{i = 0}^ n |c_i|^2 = 1
$$
and
$$
f_i({\bf k}) = \int_A 
\hat{\phi}(\hat{\bf k}, {\bf x}; \tau_3)  
\phi^*_i({\bf k}, {\bf x}) d{\bf x} = 0 
~~~~~~~~i = 0, 1,.... n -1 
\eqno(A.6)
$$
Eq. $(A.6)$ corresponds to $n$ homogeneous algebraic equations for $n + 1$
constants $c_0, c_1,...... c_n$. A choice of such $c_i$s is always possible.
Therefore
$$
\hat{\lambda}_n(\hat{\bf k}; \tau_3) \geq 
\sum_{i=0}^n |c_i|^2 \hat{\lambda}_i(\hat{\bf k};\tau_3) =
J(\hat{\phi}(\hat{\bf k}, {\bf x}; \tau_3),
 \hat{\phi}(\hat{\bf k}, {\bf x}; \tau_3))  
~~~~~~~~~~~~~~~~~~~~~~~~~~~~~~~~~~~~~~~~~~~~~~
$$
$$
~~~~~~~~~~~~~~~~~~~~~~~~~~~~~
\geq
\sum_{i=0}^{\infty} |f_i({\bf k})|^2 \lambda_i({\bf k}) =
\sum_{i=n}^{\infty} |f_i({\bf k})|^2 \lambda_i({\bf k}) \geq
\lambda_n({\bf k}) \sum_{i=n}^{\infty} |f_i({\bf k})|^2 
= \lambda_n({\bf k}) 
$$
This is Eq. (4) in the text.

\end{document}